\documentclass[epj]{svjour}	
\usepackage{epsfig}
\usepackage{amssymb}
 
\begin{document}
 
\title{Haloes and Clustering in Light, Neutron-Rich Nuclei}
 
\author{NA Orr\inst{1}}

 
\institute{Laboratoire de Physique Corpusculaire,
     IN2P3-CNRS, ISMRa et Universit\'e de Caen, F-14050 Caen cedex, France}

\date{\today}
 
\abstract{
Clustering is a relatively widespread phenomena which takes 
on many guises across the nuclear landscape.  Selected topics concerning the 
study of halo systems and clustering in light, neutron-rich nuclei are discussed 
here through illustrative examples
taken from the Be isotopic chain. }


\maketitle
 
\section{Introduction} 
 
Until relatively recently cluster studies
have been confined to the region encompassing the line of beta stability 
where the r\^ole
of $\alpha$-clustering has long been 
established \cite{Freerreview}.  
As clustering is expected to manifest itself most
strongly near thresholds \cite{Ikeda}, 
exotic structures might
be expected to form in the more weakly bound systems found near the neutron and 
proton driplines.
In practical terms the exploration of clustering in nuclei far from stability
has become technically feasible over the last 15 years through the advent of
radioactive beams and associated techniques.  
This is demonstrated most clearly by 
the discovery and
subsequent probing of the nuclear halo.

In the present paper selected topics concerning the study of halo and
molecular states in light, neutron-rich nuclei which have undergone developments
since the previous conference in this series 
are discussed.  
Following a brief review of work on $x\alpha$:Xn molecular states, emphasis
is placed on new techniques for probing correlations in two-neutron halo nuclei.
Very recent work exploring the means to produce and detect multineutron
clusters is also presented. 
As they exhibit
many of the facets of clustering and structural evolution far from stability 
the examples discussed here have been selected from the neutron-rich Be isotopes.

\section{Nuclear Molecular Clusters}

It has long been established that the $\alpha$-particle plays an important
r\^ole in the structure of light $\alpha$-conjugate (A=4n) nuclei \cite{Freerreview}.  
This
is a direct consequence of the strongly bound character of the $^{4}$He nucleus
and the weakness of the $\alpha$-$\alpha$ interaction, as evidenced by the unbound nature
of $^{8}$Be.  
Whilst an excess of
neutrons (or protons) may na\"ively be expected to dilute any underlying $\alpha$-cluster
structures, 
theoretical \cite{Sey81,AMD} and recent experimental work \cite{vonO,Fre99}
indicate that molecular type structures such as $\alpha$-chains ``bound'' by valence
nucleons may also occur.
The persistence of such cluster structures in systems lying away
from the line of  beta-stability is well illustrated, as will be discussed here, 
by the beryllium isotopes,
for which the $\alpha$-$\alpha$ system may be regarded as the basis.  

From a theoretical point of view, prescriptions such as the Molecular-Orbital Model (MO)
\cite{Sey81} or the Two-Centre Shell Model \cite{TCSM}, in which valence nucleons
are added to the single-particle orbits arising from the two-centre potential, provide a
successful and conceptually appealing framework within which to describe the properties 
of these nuclei. Moreover these orbits may be
viewed as the analogues of the $\sigma$ and $\pi$-orbitals associated with the covalent
binding of atomic molecules. The development of 
fully fledged 
Antisymmeterised Molecular Dynamics calculations (AMD), as discussed by 
Horiuchi in 
his contribution to 
this conference, is of particular 
interest as the
A-nucleon system is modelled without any a priori imposition of an underlying
cluster structure.  Recent calculations, in particular, suggest the existence of two-centred 
structures in the
Be, B and C isotopic chains with valence neutron density 
distributions exhibiting the features of molecular orbitals \cite{AMD}.

From an experimental perspective, von Oertzen \cite{vonO} has compiled systematic
evidence for the existence of dimers in $^{9-11}$Be and $^{9-11}$B.
In the case of $^{9}$Be, for example, 
the presence of a valence neutron results in a bound
(Borromean) system, the ground and excited states of which may be understood
in terms of a three-body $\alpha$:n:$\alpha$ molecular structure.  
In particular,
the rotational bands based on the ground and low-lying states exhibit large
deformations consistent with the associated molecular configurations.

\begin{figure}
\centerline{\epsfig{file=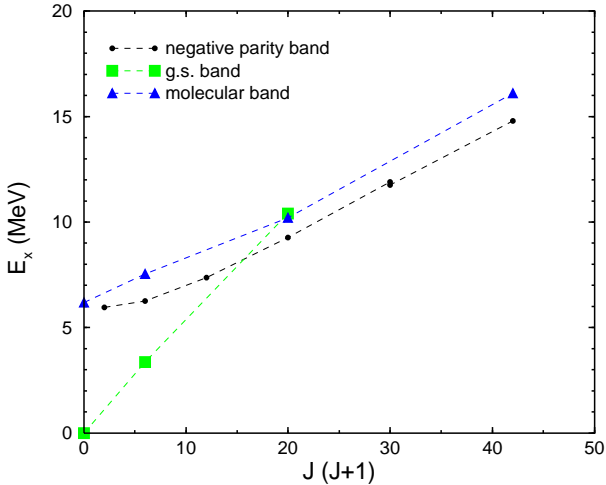,width=8.0cm}}
\caption{Spin-energy systematics for states observed in $^{10}$Be 
(from \protect\cite{Fre00}).
The trajectories for the postulated positive and negative parity molecular bands are 
indicated.}

\end{figure}

In the case of $^{10}$Be, the experimental evidence for molecular configurations
is rather less well documented.  Beyond the 
established 0$^+_2$, 2$^+_2$ and 1$^-_1$ -- 4$^-_1$ states,
the locations of the J=5 \cite{vonO} and 6 members of the of the negative 
parity band, as well as the 
J=4 and 6 members of the positive parity band have been postulated following recent
studies of the $\alpha$-$^{6}$He breakup of $^{10}$Be$^*$ \cite{Soic,Fre00}.
As displayed in Fig. 1 \cite{Fre00}, the spin-energy trajectories for the bands based 
on the
0$^+_2$ and 1$^-_1$ states at $\sim$6~MeV are consistent with 
large deformations as expected for molecular-like $\alpha$:2n:$\alpha$
structures.  Moreover, the location of the bandheads just below the threshold for
$\alpha$ + $^{6}$He decay is in accordance with the considerations of Ikeda
describing the formation of clusters \cite{Ikeda}.
Theoretical support for the postulated molecular states may be found in AMD
calculations, whereby 
well developed $\alpha$:2n:$\alpha$ configurations are predicted for the 
0$^+_2$ and 1$^-_1$ bands \cite{AMD}.  Coupled-channel calculations 
presented in these proceedings 
by Ito and
collaborators also provide support for the molecular nature of the band based on the 
0$^+_2$.  
Furthermore, extended MO model calculations \cite{Ita00} have demonstrated 
the that the 0$^+_2$ state may be well characterised by valence neutrons in
occupying the $\sigma$-orbital.

Given the existence of such molecular-type structures in $^{10}$Be, the question naturally
arises as to the existence of similar structures in more neutron-rich systems.
In this context the
dripline nucleus $^{12}$Be has been investigated via
inelastic scattering at 35~MeV/nucleon.  
Evidence was found in this measurement for the breakup into 
$^{6}$He+$^{6}$He and $\alpha$+$^{8}$He 
of states
(J=4, 6, 8)
in the excitation energy range 10-20~MeV which exhibited spin-energy systematics
characteristic of a rotational band \cite{Fre99}.
Moreover the inferred momenta of inertia ($\hbar^2/2\Im$=0.15$\pm$0.04~MeV) 
and bandhead 
energy (10.8$\pm$1.8~MeV) of the observed states
are consistent with the cluster decay
of a molecular structure which may be associated with $\alpha$:4n:$\alpha$
configurations.
As reported at this conference by Saito {\em et al.}, an experiment undertaken  
very recently at RIKEN using two neutron removal from an energetic $^{14}$Be beam has 
uncovered evidence for possible 0$^+$ and
2$^+$ states some 1.7 and 2.7~MeV above the $^6$He+$^6$He breakup threshold.  
Interestingly these
states conform well to the spin-energy systematics established in our original study.

Somewhat indirect evidence for the existence of the states observed in the two breakup experiments
may be found in the observation of highly excited states in the
$^{9}$Be($^{15}$N,$^{12}$N)$^{12}$Be reaction which are consistent with the spin-energy
systematics of a deformed rotational band \cite{Bohlen}.
The theoretical investigation of molecular configurations in $^{12}$Be represents a
somewhat more challenging venture than the lighter mass Be isotopes.  Nevertheless,
efforts are being made, as evidenced by the contribution to this conference by 
Ito and coworkers and recent papers by Itagaki {\em et al.} \cite{Ita00} and 
Descouvemont and Baye \cite{Des01}.
 
As suggested by von Oertzen \cite{vonO} and more recently by Itagaki {\em et al} 
\cite{Ita01}, 
the neutron-rich C isotopes may be expected to exhibit 3$\alpha$:Xn cluster structures.  
In this context, we have attempted to observe such states in the  
$^{12}$C($^{16}$C,$^{16}$C$^*$$\rightarrow$$^{10,12}$Be+$^{6,4}$He) reaction 
at 35~MeV/nucleon
\cite{Lea01}.  Careful analysis of the fragment coincidences could, however, only put an
upper limit of some 30~$\mu$b on the yield to states in these decay channels.  The
inability to access such states by inelastic scattering may arise from a much
smaller overlap between the $^{16}$C ground state and the cluster states than in the
case of $^{12}$Be.  In this context transfer reactions
using projectiles ($^{6,8}$He) and targets ($^{6}$Li, $^9$Be, $^{12}$C) exhibiting cluster 
structure may provide a more efficient means of accessing molecular states in neutron-rich
nuclei.

Beyond the extension of cluster studies to more exotic systems, one of the
most pressing issues is the determination of partial decay widths.  Clearly this is not a
trivial task, as evidenced by the difficulties encountered in the study of clustering
in stable nuclei.  In the case the light, neutron-rich nuclei discussed here, 
the presence
of relatively few decay channels and the more favourable signal-to-background ratio
when employing radioactive beams may, however, facilitate such measurements.

\section{Correlations in Two-Neutron Halo Nuclei}

Clustering also appears in the guise of 
neutron haloes in ground states near the 
neutron drip-line. Arguably the most intriguing are the 
Borromean two-neutron halo nuclei ($^{6}$He, $^{11}$Li and 
$^{14}$Be), in which the two-body subsystems are unbound. Such 
behaviour naturally gives rise to the question of the correlations between the 
constituents. 
Even in the case of the most studied 
of these nuclei, $^{6}$He and $^{11}$Li, little is known in this respect.
Over the last three years we have developed a number of techniques for
exploring the correlations within two-neutron halo systems \cite{FMM00,FMM01}. 

We have explored the spatial configuration of the halo
neutrons at breakup through application of the technique of intensity interferometry --
an approach first developed for stellar interferometry by Hanbury-Brown and Twiss in 
Australia in the
1950's and 60's \cite{HBT} and later extended to source size measurements in high energy
collisions \cite{Gol60}.  The principle behind the technique is as follows:  
when identical partiles are emitted 
in close proximity in space-time, the wave function
of relative motion is modified by the FSI and quantum statistical
symmetries \cite{Boa90} --- in the case of halo neutrons the overwhelming effect is that of 
the FSI \cite{FMM00}. Intensity interferometry relates this
modification to the space-time separation of the particles at emission as a 
function of the four-momenta of the particles through the correlation function 
$C_{\rm{nn}}$, which is defined as,

\begin{equation}
 C_{\rm{nn}}(p_1,p_2)=\frac{d^2n/dp_1dp_2}{(dn/dp_1)\,(dn/dp_2)} \label{e:C12}
\end{equation}

where the numerator is the measured two-particle distribution and the denominator the 
product of the independent single-particle distributions \cite{FMM00}.  
As is generally the case, the single-particle distribution
have been generated in our work via event mixing.  Importantly, in the 
case of halo neutrons
special consideration must be given to the strong residual correlations \cite{FMM00}.
Experimentally care needs to be taken to eliminate
cross talk \cite{Mar00}.

As a first step, kinematically complete measurements of breakup on a Pb target
of $^{6}$He, $^{11}$Li and 
$^{14}$Be were analysed \cite{FMM00}.  The choice of a high-Z target was made to 
privilege
Coulomb induced breakup, whereby the halo neutrons may in a first approximation 
act as spectators and for which simultaneous emission may be 
expected to occur.  The correlation functions derived from the data, assuming simultaneous
emission, were compared to
an analytical formalism based on a Gaussian source \cite{soviet}.
Neutron-neutron separations of $r_{nn}^{RMS}=5.9\pm1.2$~fm ($^{6}$He), 
$6.6\pm1.5$~fm ($^{11}$Li) and $5.6\pm1.0$~fm ($^{14}$Be) were thus extracted.  These 
results appear to
preclude any strong dineutron component in the halo wavefunctions at breakup.
It is interesting in this context to compare these results to the RMS neutron-proton
separation of 3.8~fm in the deuteron (the only bound two nucleon system).

The same analysis has been applied to dissociation of $^{14}$Be by a C target,
in order to investigate the influence of the reaction mechanism.  A 
result which hints at a somewhat larger separation, $r_{nn}^{rms}=7.6\pm1.7$~fm, was 
obtained. This raises the question as to whether simultaneous emission can be
assumed a priori. In principle, the analysis of the correlation function in two
dimensions, transverse and parallel to the total momentum of the pair, would
allow for the unfolding of the source size and lifetime \cite{soviet}. Such an
analysis requires a much larger data set than presently available. 
The two-neutron halo, however, is far less complex than
the systems usually studied via interferometry (for example, heavy-ion 
collisions \cite{Boa90}).
Moreover, the simple three-body nature of the system breaking up suggests
that any delay in the emission of one of the neutrons will arise
from core-n FSI/resonances in the exit channel, a process that may be expected to be
enhanced for nuclear induced breakup.

Correlations in three-particle decays are commonly encountered in particle
physics and are typically analysed using plots of the
squared invariant masses of particle pairs $(M_{ij}^2,M_{ik}^2)$, with
$M_{ij}^2=(p_i+p_j)^2$; a technique developed by the Australian physicist Richard Dalitz
in the early 1950's \cite{Dal53}. In these Dalitz plot representations, 
FSI or resonances lead to a
non-uniform population of the surface within the kinematic boundary defined by
energy-momentum conservation and the decay energy. 
In the present case, the core+n+n system exhibits a distribution of decay
energies ($E_{\rm{d}}$). The $E_{\rm{d}}$ associated 
with each event
will thus lead to a different kinematic boundary, and the resulting plot
containing all events cannot be easily interpreted. We have thus introduced a
normalised invariant mass, 

\begin{equation}
 m_{ij}^2 = \frac{M_{ij}^2-(m_i+m_j)^2}{(m_i+m_j+E_{\rm{d}})^2-(m_i+m_j)^2}
\end{equation}

which ranges between 0 and 1 
(that is, a relative energy $E_{ij}=M_{ij}-m_i-m_j$ 
between 
0 and $E_{\rm{d}}$) 
for all events and exhibits a single kinematic boundary. 
Examples of how n-n and core-n FSI may manifest themselves 
in the Dalitz plot for the decay of $^{14}$Be are illustrated in Fig.~2, whereby 
events have been simulated according to the simple interacting phase-space
model described in ref. \cite{FMM01}. The inputs were an 
$E_{\rm{d}}$ distribution following that measured \cite{Lab01}, the 
$C_{\rm{nn}}$ obtained with the C target, and a
core-n resonances with $\Gamma=0.3$~MeV at $E_0=0.8$~MeV.  Note that due to the
normalisation the (squared) core-neutron invariant mass does not present 
a simple structure directly
related to the energy of the resonance/FSI \cite{FMM01}. 

\begin{figure}
\centerline{\epsfig{file=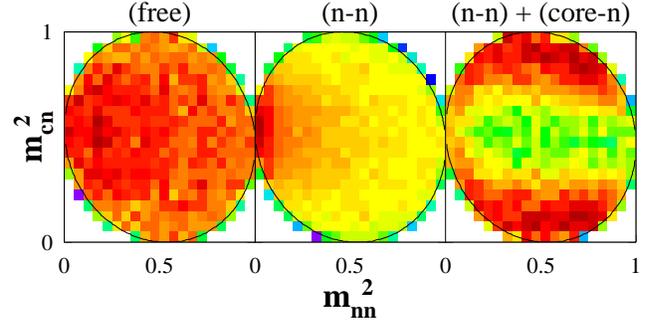,width=8.5cm}}
\caption{Dalitz plot for the simulated decay (see text) of $^{14}$Be.  
In the left panel no FSI are included.}
\end{figure}

As described in ref. \cite{FMM01}, the data from the dissociation of $^{14}$Be on C and Pb
targets have been analysed in terms of Dalitz plots and compared to the interacting 
phase-space calculations.  The results for breakup on the Pb target are shown in Fig 3.
The good agreement obtained through the inclusion of only the n-n FSI suggests that, within
the presently available statistics,
the two-neutrons are emitted essentially simultaneously.  

Despite much reduced statistics for the triple coincidence events needed for this 
analysis, breakup on the C target is consistent with a finite delay between the emission of
the two neutrons \cite{FMM01} . This may be 
related to the population of core-neutron resonances
as suggested by more classical analyses of $^{12}$Be-neutron
events \cite{Orr00}.  By fixing the n-n source size to that derived from the
$C_{nn}$ for the Pb target, an average delay 
in the emission for breakup by C of $150^{+250}_{-150}$~fm/$c$ was estimated.

Finally, in the context of the influence of the reaction mechanism, it is worthwhile 
noting that whilst some 35\% of the two-neutron removal 
cross section on the Pb target is  
attributable to nuclear 
induced breakup \cite{Lab01}, the requirement of two neutrons in coincidence with the
$^{12}$Be core in the present analysis reduces this to some 15\% -- approximately half
of the two-neutron removal cross section arises from absorption.

\begin{figure}
\centerline{\epsfig{file=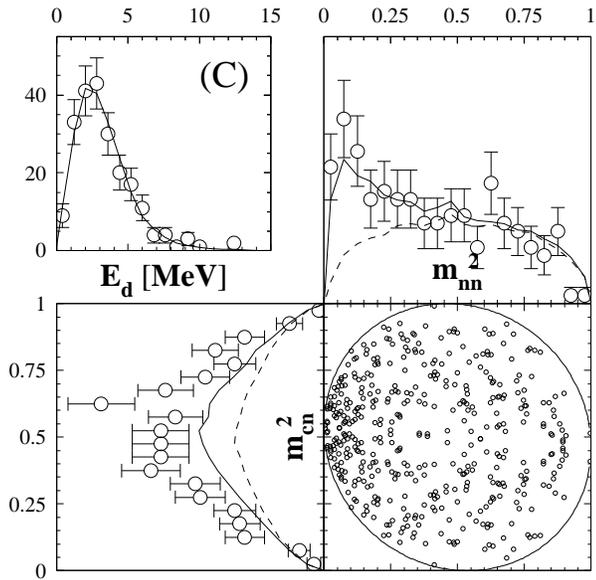,width=8.0cm}}
\caption{Dalitz plot and the projections onto the squared invariant masses
for the dissociation of $^{14}$Be by Pb. The lines are the 
phase-space model simulations
with/without (solid/dashed) n-n FSI. The inset shows the measured $E_d$ spectrum.}
\end{figure}

\section{Multineutron Clusters}

It is interesting to speculate that multineutron halo nuclei
and other very neutron-rich systems may contain components 
of the wavefunction in which
the neutrons present a relatively compact cluster-like configuration.  If this
were to be the case, then the dissociation of beams of such nuclei may offer a means
to produce bound multineutron clusters (if they exist) and, more generally
study multineutron correlations.  

To date the majority of searches for 
multineutron systems have relied on very low (typically $\sim$1~nb) cross 
section double-pion 
charge exchange (D$\pi$CX) and heavy-ion transfer
reactions (see, for example, refs \cite{pion,HI}).  In the case of dissociation of
an energetic beam of a very neutron-rich nucleus, relatively high cross sections
(typically $\sim$100~mb) are encountered.  Thus, even only a small component
of the wavefunction corresponding to a multineutron cluster could result in a measurable
yield with a moderate secondary beam intensity.  Furthermore the backgrounds
encountered in D$\pi$CX and heavy-ion transfer
reactions are obviated in direct breakup.

The difficulty in this approach lies in the direct detection of a $^A$n cluster.
One avenue we are exploring is to use the detection of the recoiling
proton in a liquid scintillator \cite{FMM01a}.  The advantage 
of a liquid scintallator is that neutrons may be discriminated from
the $\gamma$ and cosmic-ray backgrounds. 
Careful source and cosmic-ray calibrations 
permit the charge deposited and hence the energy ($E_p$) of the recoiling proton  
to be determined.
This may be compared to the energy derived from the measured time-of-flight
($E_n$):
for a single neutron and an ideal detector, $E_p$/$E_n$$\leq$1; for a 
realistic detector with 
finite resolution the limit is $\sim$1.4.  In the case of a 
multineutron cluster ($^A$n) $E_p$ can exceed the incident energy per nucleon
and
$E_p$/$E_n$ will take on a range of values 
extending beyond 1.2 --- up to $\sim$3 in the case of A=4 (Fig. 4).

\begin{figure}
\centerline{\epsfig{file=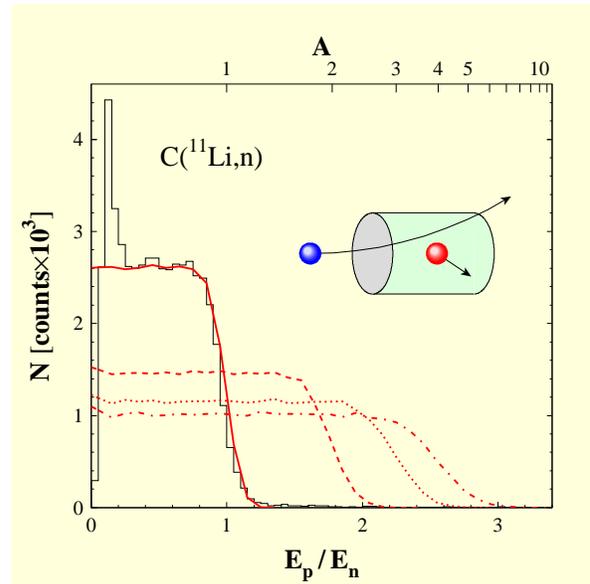,width=7.75cm}}
\caption{$E_p/E_n$ for A=1 (solid line), 2 (dashed), 3 (dotted) and 4 (dot-dashed).
In the case of A=1, comparison is made to single neutron events from $^{11}$Li + C. }
\end{figure}  

Analyses of data already available from the breakup of 
$^{14}$Be by C show some 7 events with $E_p/E_n$>1.4 (Fig. 5).  Furthermore all 7 events
are in coincidence with $^{10}$Be fragments.  Intriguingly, preliminary estimates suggest
that the detection of two or more
neutrons in a single module (the most likely source of background) should produce 
much less that one event with $E_p/E_n$>1.4 in coincidence with $^{10}$Be.  Further 
detailed analysis 
of these events and all possible sources of background is presently underway.
In addition, higher statistics measurements employing beams such as $^{8}$He are
currently being planned.

\begin{figure}
\centerline{\epsfig{file=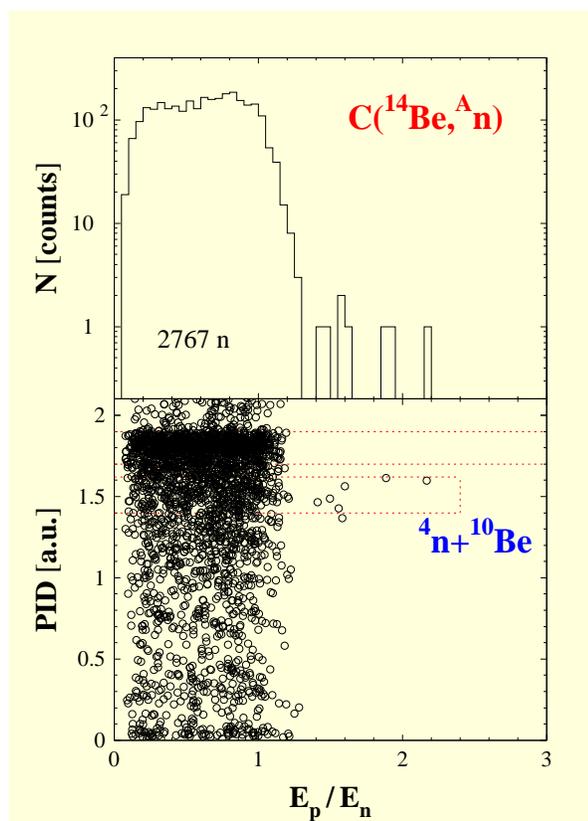,width=7.75cm}}
\caption{Particle identification (PID) versus $E_p/E_n$ for the breakup of $^{14}$Be by C.  
The two horizontal
bands deliniated by the dotted lines correspond to the $^{12}$Be (upper)
and $^{10}$Be(lower) fragments. }
\end{figure}

\acknowledgement

It is a pleasure to thank the members of the E281 and E295 collaborations and, in
particular,  
the DEMON and CHARISSA crews. I would also
like to draw attention to the key r\^oles played
by Martin Freer (Be cluster studies), Miguel Marqu\'es (correlations and neutron clusters) 
and Marc Labiche ($^{14}$Be breakup).
Finally, the support provided by the staffs of LPC and GANIL in preparing and executing
the experiments is gratefully acknowledged. 

This work was funded under the
auspices of the IN2P3-CNRS (France) and EPSRC (United Kingdom). Additional
support from the ALLIANCE programme (Minist\`ere des Affaires Etrang\`eres and
British Council) and the Human Capital and Mobility Programme of the European
Community (Access to Large Scale Facilities) is also acknowledged.

\end{document}